\documentclass[aps,prl,reprint,superscriptaddress,showpacs]{revtex4-1}
\usepackage[latin9]{inputenc}
\usepackage{amsmath}
\usepackage{graphicx}
\usepackage{subscript}

\makeatletter

\usepackage{epstopdf}

\makeatother

\begin{document}

\title{Magnetoresistance and Quantum Oscillations of an Electrostatically-Tuned
Semimetal-to-Metal Transition in Ultra-Thin WTe\textsubscript{2}}

\author{Valla Fatemi}

\affiliation{Department of Physics, Massachusetts Institute of Technology, Cambridge,
Massachusetts 02139, USA}

\author{Quinn D. Gibson}

\affiliation{Department of Chemistry, Princeton University, Princeton, NJ 08544,
USA}

\author{Kenji Watanabe}

\affiliation{Advanced Materials Laboratory, National Institute for Materials Science,
1-1 Namiki, Tsukuba 305-0044, Japan}

\author{Takashi Taniguchi}

\affiliation{Advanced Materials Laboratory, National Institute for Materials Science,
1-1 Namiki, Tsukuba 305-0044, Japan}

\author{Robert J. Cava}

\affiliation{Department of Chemistry, Princeton University, Princeton, NJ 08544,
USA}

\author{Pablo Jarillo-Herrero}

\affiliation{Department of Physics, Massachusetts Institute of Technology, Cambridge,
Massachusetts 02139, USA}

\date{\today}
\begin{abstract}
We report on electronic transport measurements of electrostatically
gated nano-devices of the semimetal WTe\textsubscript{2}. High mobility
metallic behavior is achieved in the 2D limit by encapsulating thin
flakes in an inert atmosphere. At low temperatures, we find that a
large magnetoresistance can be turned on and off by electrostatically
doping the system between a semimetallic state and an electron-only
metallic state, respectively. We confirm the nature of the two regimes
by analyzing the magnetoresistance and Hall effect with a two-carrier
model, as well as by analysis of Shubnikov-de Haas oscillations, both
of which indicate depletion of hole carriers via the electrostatic
gate. This confirms that semiclassical transport of two oppositely
charged carriers accurately describes the exceptional magnetoresistance
observed in this material. Finally, we also find that the magnetoresistance power law is sub-quadratic and density-independent, suggesting new physics specifically in the semimetallic regime.
\end{abstract}


\maketitle

Semimetals, electronic systems with partially populated bands of both
positive and negative curvature, have been sources of novel physics
since at least 1930, with the discovery of Shubnikov-de Haas (SdH)
oscillations in bismuth\citep{schubnikow1930magnetische,shoenberg2009magnetic}.
The semimetal graphite, when thinned down to the electronically two-dimensional
(2D) limit, displays remarkable physics that continues to be explored\citep{goerbig2011electronic,dassarma2011electronic}.
Even more recently, three-dimensional (3D) bulk semimetals have had
a resurgence of interest due to recently discovered experimental and
theoretical behaviors, including unusually large magnetoresistances
\citep{ali2014largenonsaturating,shekhar2015extremely} and new topological
electronic states known as Dirac and Weyl semimetals \citep{vafek2014diracfermions,xu2015discovery}. 

The compound WTe\textsubscript{2} bridges both of these phenomena:
experimentally it showcases exceptionally large, quadratic magnetoresistance
in magnetic fields up to 60T \citep{ali2015correlation}, and theoretically it is predicted to be a Weyl semimetal in 3D \citep{soluyanov2015typeiiweyl}
and near a quantum spin Hall state in 2D \citep{qian2014quantum}.
The observed magnetoresistance in bulk samples has been proposed to
be the result of near-compensated electron and hole carriers, described
by a semiclassical two carrier model \citep{ali2014largenonsaturating,babkin1971influence,alekseev2015magnetoresistance}.
However, a number of questions have been raised regarding the mechanisms
at play. First, quantum oscillations revealed multiple Fermi pockets
as well as imperfect compensation \citep{zhu2015quantum,pletikosic2014electronic,wu2015temperatureinduced},
suggesting other effects may play a role. Moreover, the small Fermi
energy and orbital helicity of the Fermi surface suggest that the
band structure and scattering mechanisms, respectively, are liable
to change under application of a magnetic field \citep{jiang2015signature,rhodes2015roleof,das2016layerdependent}.
To elucidate the origins of the magnetoresistance, a natural experiment
would be to study the effect of changing the carrier density of a
thin sample in-situ, which recent transport studies have attempted with varying results \citep{wang2016breakdown,wang2015tuningmagnetotransport,wang_direct_2016}.
A major difficulty at the ultrathin limit is that sample quality degrades significantly as
the thickness is reduced: oxidation induces insulating behavior below
6 layers thick, rendering phenomena related to the high quality bulk
crystals inaccessible in the 2D limit \citep{wang2015tuningmagnetotransport,lee2015tungsten,ye_environmental_2016}.

In this Letter, we investigate exfoliated WTe\textsubscript{2} devices
that are fabricated in a glove box under inert atmosphere in order
to minimize degradation\citep{fei_topological_2016}. Doing so enables creation of high quality
nanodevices that display the intrisic physics of the material \citep{tsen2015structure}.
At low temperatures, we use an electrostatic gate to dope the system
from a semimetallic regime to a single-carrier regime. During this crossover
we observe an evolution of the magnetoresistance and Hall effect that is qualitatively
well captured by a semiclassical two-carrier model. We additionally
observe that the exponent of the magnetoresistance power law is subquadratic and gate-independent within the semimetallic regime. Finally, the semimetal-to-metal
transition is further confirmed by analysis of SdH oscillations, which
give insight to the low-energy band structure.

To fabricate the devices, WTe\textsubscript{2} is exfoliated in an
argon atmosphere ($<1$ ppm oxygen) and then encapsulated between
layers of hexagonal Boron Nitride (h-BN) with a polymer pick-up and
transfer technique\citep{dean2010boronnitride}. For the device discussed
in the main text, we also include few-layer graphene (FLG) between
the WTe\textsubscript{2} and the bottom h-BN layer as an electronic
contact. This FLG layer has a natural lateral gap between two independent
sheets, which the WTe\textsubscript{2} spans. Finally, we contact
the FLG with evaporated Cr-Au via the edge-contact technique\citep{wang2013onedimensional},
and then etch the device into a Hall bar geometry with a reactive
ion etch. The FLG serves as an intermediary conductor between the
evaporated electrodes and the WTe\textsubscript{2} while also maintaining
a good van der Waals seal with the encapsulating h-BN\citep{cui2015multiterminal}.
The device sits on a doped silicon wafer with $285nm$ of thermal
SiO\textsubscript{2}, which serve as the back-gate electrode and dielectric, respectively.
All electronic transport measurements are conducted in a four-probe
configuration. An optical micrograph and a schematic cross-section
of the device are shown in Fig. 1(a-b). Here we report on the behavior
of a 3-layer thick device (sample A, thickness confirmed by AFM and
Raman spectroscopy \citep{wang2015tuningmagnetotransport}), and in
the supplement we show data on additional devices \citep{SI}.

\begin{figure}[t]
\includegraphics{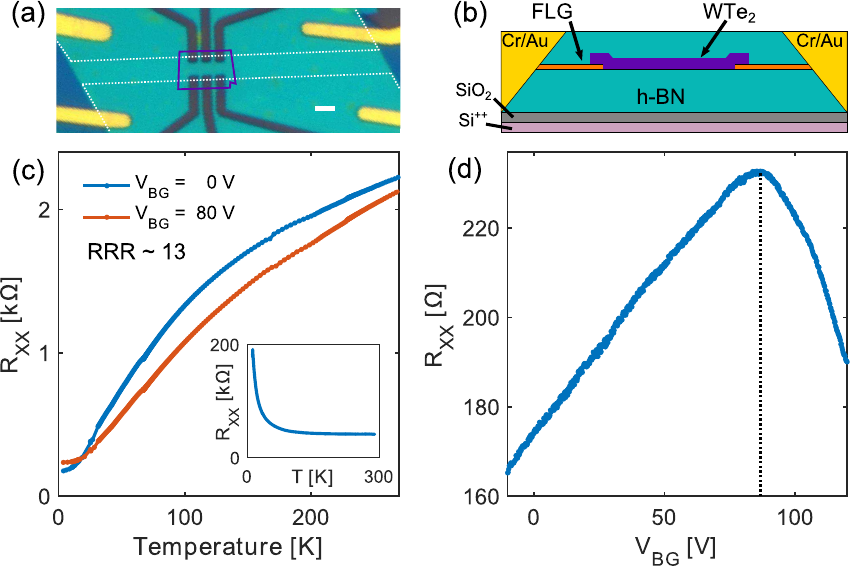}
\caption{\label{fig:1} (Color online) (a) Optical image of a
completed device. The edges of the original FLG flakes are indicated
by dashed white line, and the original boundary of the WTe\protect\textsubscript{2}
flake is outlined in purple. The dark regions are where the stack
was fully etched to the SiO\textsubscript{2} substrate. Solid white scale bar is 2 microns long. (b) Cross-sectional schematic of the device structure,
with all components labeled. (c) Temperature dependence of device A
at two representative gate voltages. The average RRR for all gate
voltages is 13. Inset: Temperature dependence of a similar-thickness
WTe\protect\textsubscript{2} device fabricated in air and without
encapsulation. (d) Gate dependence of $R_{xx}$ at $B=0T$ and $T=4K$. The
vertical dashed line at $V_{BG}=87$ indicating the resistance maximum.}
\end{figure}

The first indication that inert atmosphere fabrication maintains crystal integrity is the temperature dependence of the resistance, shown in Fig. 1(c). We observe metallic behavior
with a residual resistivity ratio (RRR) of 13. In contrast, non-encapsulated
devices fabricated in air display insulating behavior
(inset of Fig. 1(c)). In fact, in the literature a RRR of 13 is only achieved for samples greater than 33nm thick\protect\citep{wang2015tuningmagnetotransport}. 
Applying a bias $V_{BG}$ to the electrostatic gate, we capacitively modulate
the carrier density in the sample, finding that the low-temperature
resistivity increases linearly with $V_{BG}$ until $\sim87$V, beyond which
the resistivity drops sharply (see Fig. 1(d))\protect\citep{cooldownnote}. The noted metallic temperature dependence with RRR of order 10 is observed for all gate voltages, with two representative gate voltages shown in Fig. 1(c).

\begin{figure}[t]
\includegraphics{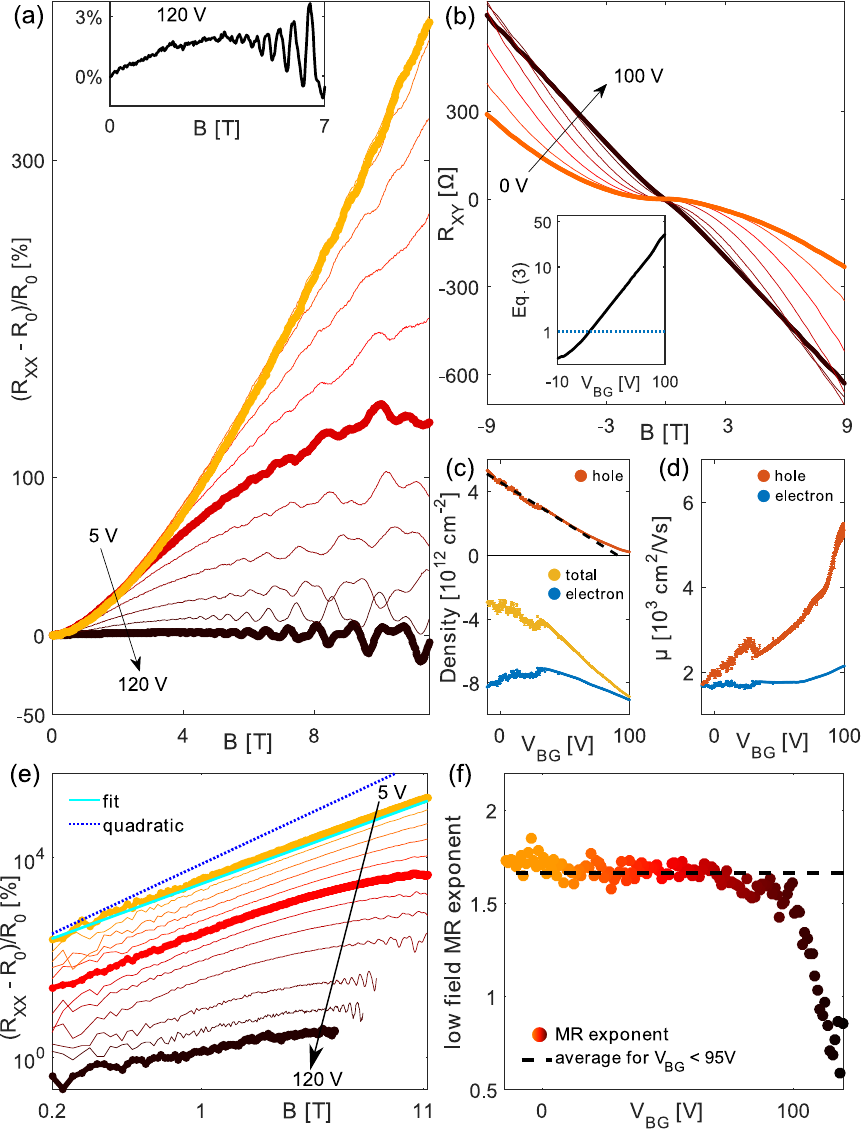}

\caption{\label{fig:2} (Color online) (a) MR as a function of magnetic field
for a range of $V_{BG}$ at $T= 30 mK$. Inset: a zoom-in of the data at $V_{BG}=120V$. (b) $R_{xy}$ as a function of magnetic
field for a range of $V_{BG}$ at $T= 300 mK$, displaying a transition from non-linear
to linear Hall effect \protect\citep{cooldownnote}. Inset: Extracted saturation
parameter (equation \eqref{eq:satparam}) as a function of gate voltage
(black) at $B=10T$. The blue dotted line is unity. (c) Individual carrier densities
and the total carrier density, including charge sign, from the semiclassical
model fit. Black dashed line is an extrapolation of the hole density
from the slope at lower gate voltages. (d) Mobilities from the
model fit. Error bars in (c-d) represent 95\% confidence intervals of the fit. (e) Same data as (a) in log-log format with each curve
offset vertically for clarity. The blue dotted line is a quadratic
power law, whereas the cyan line is a power law fit to the data at
$V_{BG}=10V$. (f) The fitted exponent of the MR power law for $B\in\left[0.4,1.5\right]T$
as a function of gate voltage. The limits were chosen to avoid
the high relative noise at low fields ($B<0.2T$) and onset of saturation
at high field. Dot color corresponds to gate voltage in accordance with (a) and (e). Dot size is larger than the 95\% confidence interval of the fit. The black dashed line indicates the mean exponent for $V_{BG}<95V$. }
\end{figure}

We then measure both the longitudinal ($R_{xx}$) and transverse ($R_{xy}$)
resistances as a function of magnetic field and gate voltage in order
to investigate the magnetoresistance behavior at different total carrier
densities. Both measurements are shown in Fig. \ref{fig:2}(a-b) for
a range of gate voltages, where in Fig \ref{fig:2}(a) we plot the
magnetoresistance ratio (MR), defined as $\left(R_{xx}(B)-R_{xx}(0)\right)/R_{xx}(0)$
\protect\citep{cooldownnote}. The MR shows a clear transition from strong,
superlinear behavior near zero gate voltage ($\sim400\%$ increase
at 11.5T) to suppressed MR ($<3\%$) at the highest gate voltage (inset
of Fig. 2(a)). During the crossover at intermediate gate voltages,
we find that $R_{xx}$ saturates at large $B$, as expected for a non-compensated
semimetal \protect\citep{du2005metalinsulatorlike,alers1953themagnetoresistance}.
Importantly, we observe that this crossover coincides with the Hall
effect transitioning from nonlinear to linear magnetic field dependence,
indicating a transition from two carrier types to a single carrier
type.

To investigate this behavior in more detail, we employ the semiclassical
two-carrier model, which gives the following equations for the longitudinal
and transverse resistivity of a semimetal:
\begin{eqnarray}
\rho_{xx} & = & \frac{1}{e}\frac{n\mu_{n}+p\mu_{p}+\left(n\mu_{p}+p\mu_{n}\right)\mu_{n}\mu_{p}B^{2}}{\left(n\mu_{n}+p\mu_{p}\right)^{2}+\left(p-n\right)^{2}\mu_{n}^{2}\mu_{p}^{2}B^{2}}\label{eq:rxx}\\
\rho_{xy} & = & \frac{1}{e}\frac{\left(p\mu_{p}^{2}-n\mu_{n}^{2}\right)B+\left(p-n\right)\mu_{n}^{2}\mu_{p}^{2}B^{3}}{\left(n\mu_{n}+p\mu_{p}\right)^{2}+\left(p-n\right)^{2}\mu_{n}^{2}\mu_{p}^{2}B^{2}},\label{eq:rxy}
\end{eqnarray}
which include the electron charge $e$ and four free parameters: the
density ($n$, $p$) and mobility ($\mu_{n,p}$) of each carrier type,
where $n$ and $p$ refer to electron-like and hole-like carriers,
respectively. We fit $R_{xx}(B)$ and $R_{xy}(B)$ simultaneously
to extract all four quantities at each gate voltage. (Here we constrain
the electron density based on the SdH analysis presented below, but
the important qualitative behavior doesn't require this. See SI for
the details\protect\citep{SI}.) The fit parameters are plotted as a function
of $V_{BG}$ in Fig. 2(c-d). Most notably, we find that the hole density
decreases monotonically to nearly zero at $V_{BG}\sim100V$, consistent with
our earlier, qualitative interpretation of the data. 

We highlight now a particular aspect of the semiclassical model regarding
the onset of saturation (or absence thereof). The condition for onset
of significant saturation in the MR is given by the ratio of the two
terms in the denominator of equation \eqref{eq:rxx}: 
\begin{equation}
\frac{(p-n)^{2}\mu_{n}^{2}\mu_{p}^{2}B^{2}}{\left(n\mu_{n}+p\mu_{p}\right)^{2}}\sim1,\label{eq:satparam}
\end{equation}
which is plotted in the inset to Fig. 2(b) for $B=10T$. This condition
is satisfied for $V_{BG}>22V$, around which gate voltage we see the
onset of saturation in the MR. The ratio is greater than 10 for $V_{BG}>70V$
for which gate voltages we observe near complete saturation at high
magnetic field. When seemingly unsaturating MR is observed in a semimetal,
equation \eqref{eq:satparam} can set a bound for the degree of non-compensation.
For example, Ali, et al, \protect\citep{ali2014largenonsaturating} measure
effectively no saturation in bulk WTe\textsubscript{2} in fields
up to 60T. From this and the mobility (which sets the magnitude of
the MR), one may estimate an upper bound for the relative carrier
density difference $\left|\frac{n-p}{n+p}\right|\lesssim0.3\%$, which
is an order of magnitude smaller than that estimated from SdH measurements
of the bulk Fermi surface\protect\citep{zhu2015quantum,rhodes2015roleof}.
This suggests that other factors may be involved in avoiding saturation
at the highest fields for 3D samples \protect\citep{wang2016breakdown}. A
possible avenue would be to investigate the role of mobility anisotropies,
known to be important in bismuth\protect\citep{collaudin2015angledependence}.
Nonetheless, for the range of magnetic fields studied here, our analysis
confirms that the large MR observed in WTe\textsubscript{2} is rooted
in the near-compensation of electron and hole densities.

Upon closer examination of our data, we find that the MR is actually
sub-quadratic\footnote{While the non-parabolicity of the low-field MR may introduce
a systematic error to the fits by the semiclassical MR equations,
it does not change the qualitative behavior of the fit parameters
nor our general conclusions \textendash{} similar behavior is observed
when fitting the Hall data with a constraint given only by the zero-field
longitudinal resistance (see SI for details\protect\citep{SI}). Lacking a physical model for the modified power law, we choose not to artificially alter the two-carrier model for the fits.}.
We can see this clearly by looking at the data in log-log format,
as shown in Figure 2(e). An explicit fit of the power law gives an
exponent near $1.6$, which is consistent for sample A in two different cooldowns as well as for additional devices presented in the supplement\protect\citep{SI}.
The exponent is stable for nearly all gate voltages for all devices, except above
$V_{BG}\sim100V$ for sample A, beyond which the exponent drops rapidly (Fig. 2(f)).
This drop-off coincides with the indications of depletion of
the holes, reinforcing that the MR can be understood as that of a
standard semimetal with a correction to the exponent.

The origin of the correction to the exponent is as yet unclear. Many studies on bulk WTe\protect\textsubscript{2} have noted a large, quadratic MR \protect\citep{ali2014largenonsaturating,wu2015temperatureinduced,rhodes2015roleof,zhu2015quantum,wang2015tuningmagnetotransport,wang_direct_2016,cai2015drastic}, but some that explicitly fit the power law also find sub-quadratic exponents \cite{thoutam_temp_2015,wang2016breakdown}. While we observe a similar deviation, our sample is in the 2D limit, whereas the Fermi surface of bulk WTe\protect\textsubscript{2} is 3D in character\protect\citep{zhu2015quantum,rhodes2015roleof}, suggesting that dimensionality is not a driving factor. Additionally, we find that the power law is density-independent in the semimetallic regime, and almost no MR is observed in the electron-only regime. This suggests that the coexistence of electrons and holes is crucial. One might look to semimetallic Boltzmann transport models which predict that boundary effects can generate linear MR when the valley recombination length is comparable to the transport channel width \protect\citep{babkin1971influence,alekseev2015magnetoresistance}, but this requires classically strong magnetic fields not present here ($\mu_{n,p}B\lesssim3$ for this work).
Inter-species drag effects may also play a role in magnetoresistance \protect\citep{coulombdragreview}, but finite inter-valley scattering should suppress such physics. The question of the unusual magnetoresistance poses an experimental and theoretical challenge for future investigation.

\begin{figure}[t]
\includegraphics{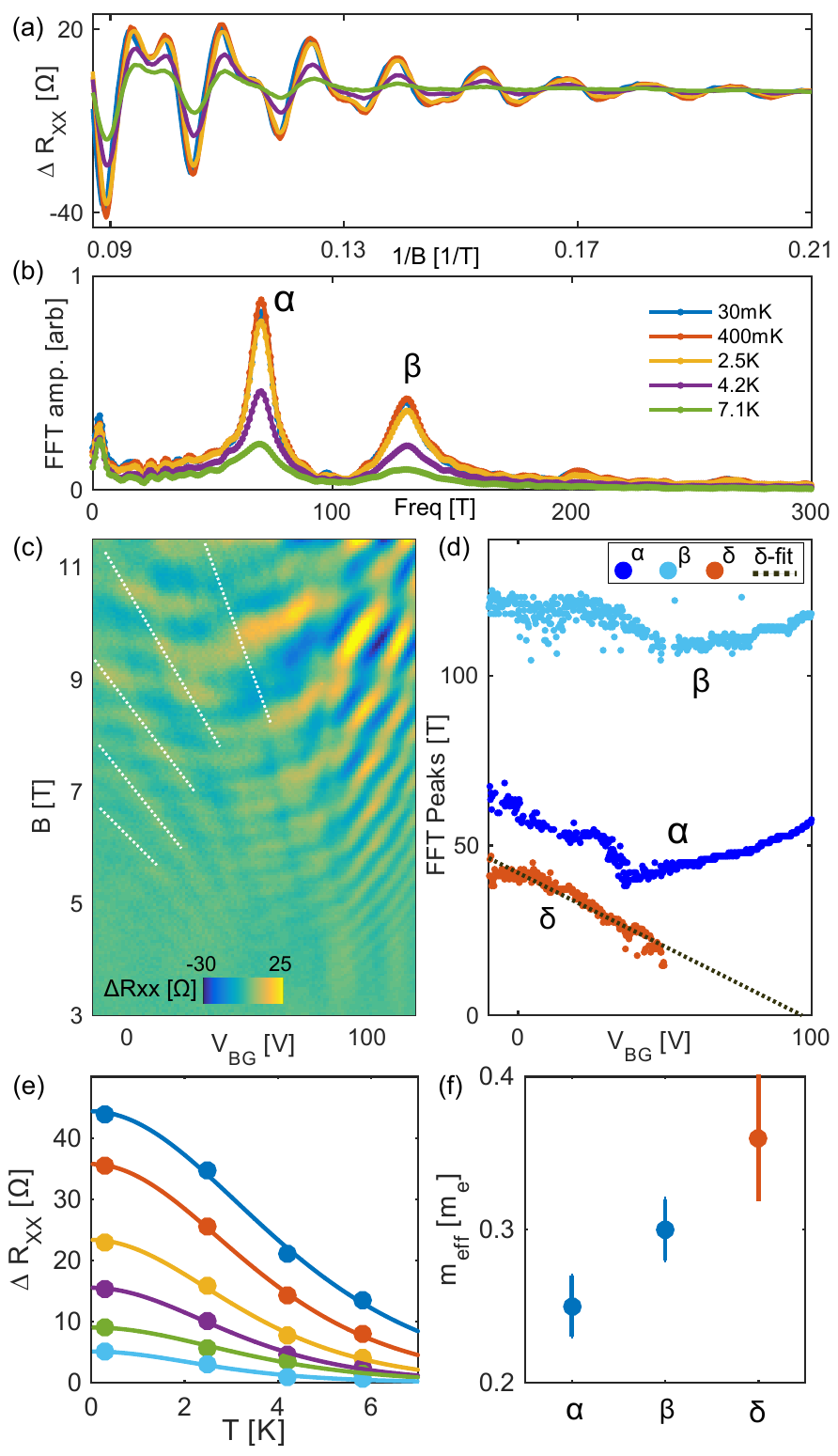}
\caption{\label{fig:3} (Color online) (a) SdH oscillations in $\Delta R_{xx}(1/B)$
at different temperatures and (b) their fast Fourier transforms (FFTs) at $V_{BG}=120V$.(c) Gate- and field-dependence of $\Delta R_{xx}$ at $T=30mK$. White dotted lines
guide the eye to the $\delta$ pocket oscillation. (d) FFT peak frequencies
(dots) at each gate voltage, and extrapolation of the $\delta$ frequency
to zero (dotted line). (e) Temperature dependence of $\Delta R_{xx}$
maxima associated with the $\alpha$ pocket at $V_{BG}=120V$. (f)
Effective mass of each pocket, from fits such as in (e) (see SI for
more details\protect\citep{SI}).}
\end{figure}

We now turn to analysis of the quantum oscillations. In Fig. \ref{fig:3}(a)
we show representative measurements of $\Delta R_{xx}(1/B)$, the
resistivity after subtracting a smooth quadratic background above 2 T, in the
electron-only regime ($V_{BG}=120V$). At this gate voltage, two oscillations are clearly
visible, confirmed by the two peaks visible in the Fourier transform
of the data in Fig. \ref{fig:3}(b), which we label $\alpha$ and
$\beta$. In Fig. \ref{fig:3}(c) we show a map of $\Delta R_{xx}$
with respect to both $B$ and $V_{BG}$. A third, hole-like dispersing
oscillation is additionally visible for lower gate voltages (highlighted
by dotted white lines), which we label $\delta$. This is made clearer
by fast Fourier transform (FFT) analysis, as shown in Figure 3(d) where the three
observed peak frequencies are tracked as a function of gate voltage\protect\citep{SI}.
The $\alpha$ and $\beta$ frequencies disperse weakly with gate voltage
in a very similar manner as the electron carrier density from the
semiclassical analysis. The non-monotonic gate dependence of the electron
bands is likely the result of strong electric field effects, which
can be explored in future devices employing a dual-gate geometry\protect\citep{qian2014quantum,taychatanapat2010electronic}.
The third oscillation frequency, $\delta$, decreases monotonically
with increasing gate voltage, which we can extrapolate to zero frequency
(full depletion) at roughly $V_{BG}=90V$. This depletion voltage
agrees with what is found for the positive charge carriers in the
semiclassical analysis, so we ascribe the $\delta$ pocket to a
valence band. Curiously, this
depletion also coincides with a peak in the amplitude of the SdH oscillations
of the $\alpha$ and $\beta$ pockets (see Fig. \ref{fig:3}(c)),
suggesting that scattering between the electron and hole valleys is
important. 

We also conduct temperature dependence of the SdH oscillations. Fitting the Lifshitz-Kosevich formula to the temperature dependence of resistance oscillation maxima  allows for extraction of the effective mass. In multi-band systems care must be taken to extract the oscillation amplitude of an individual frequency, which can be done by an appropriate FFT analysis (see SI for more details \protect\citep{SI}). An
example fit for the $\alpha$ pocket oscillation is shown in Fig. 3(e), giving $m_{\alpha}/m_{e}=0.25\pm0.02$,
where $m_{e}$ is the bare electron mass. Similar analysis for the
other bands give $m_{\beta}/m_{e}=0.30\pm0.02$ and $m_{\delta}/m_{e}=0.36\pm0.04$
(see Fig. 3(f))\protect\citep{SI}\footnote{Fits to the temperature dependence of multiple resistance maxima at several (nearby) gate voltages are averaged to obtain the mean effective mass. At least 20 different fits are made for each Fermi pocket.}. These values are somewhat smaller
than those reported for bulk crystals, which range from 0.3 to 1
$m_{e}$ \protect\citep{cai2015drastic,zhu2015quantum}. The electron effective
masses are in good agreement with theory for the monolayer, while
the hole mass is again smaller than that prediction by about a factor
of two\protect\citep{lv2015perfect}. We note that explicit predictions for
few-layer WTe\textsubscript{2} have not yet been made. 

In summary, we investigated electronic transport in encapsulated,
ultra-thin WTe\textsubscript{2}. We find that a strong, intrinsic
MR can be turned off by electrostatically doping the sample from a
semimetallic state to an electron-only regime. We confirm the basis
of the two regimes by simultaneously analyzing the MR
and the Hall effect with a two-carrier model as well as by analysis
of Shubnikov-de Haas oscillations, both of which indicate depletion
of hole-like carriers in the suppressed-MR regime. These observations
confirm that the MR in WTe\textsubscript{2} is qualitatively explained
by semiclassical transport of a semimetal, with potentially
new physics in a modified exponent to the MR power law.%

\begin{acknowledgments}
This work was partly supported by the DOE, Basic Energy Sciences Office, Division of Materials Sciences and Engineering, under award DE-SC0006418 (sample fabrication and measurements) and partly through AFOSR Grant No. FA9550-16-1-0382 (data analysis), as well as the Gordon and Betty Moore Foundation's EPiQS Initiative through Grant No. GBMF4541 to P.J.H.. Device nanofabrication was partly supported by the Center for Excitonics, an Energy Frontier Research Center funded by the DOE, Basic Energy Sciences Office, under Award No. DE-SC0001088. Crystal growth at Princeton University was supported by the NSF MRSEC program grant DMR-1005438. This work made use of the Materials Research Science and Engineering Center Shared Experimental Facilities supported by NSF under award DMR-0819762. Sample fabrication was performed partly at the Harvard Center for Nanoscale Science supported by the NSF under grant no. ECS-0335765. We thank E. Navarro-Moratalla, J. D. Sanchez-Yamagishi, and L. Bretheau for discussions and Sanfeng Wu for help with crystal exfoliation.
\end{acknowledgments}

\end{document}